\documentclass[aps,prb,twocolumn,doublespace,showpacs,superscriptaddress,groupedaddress,floatfix]{revtex4-1}
\usepackage{graphicx}
\usepackage{color}
\usepackage{amsmath}
\usepackage{amssymb}
\usepackage{comment}
\usepackage{gensymb}
\usepackage{longtable}
\usepackage{setspace}
\usepackage{tikz}
\usetikzlibrary{shapes,arrows}
\usepackage{titlesec}

\definecolor{scarred}{rgb}{0.75,0.0,0.0}
\begin{document}
\title{Weak ferromagnetism and magnetization reversal in YFe$_{1-x}$Cr$_x$O$_3$}
\author{Nagamalleswararao Dasari$^{1}$}\email{nagamalleswararao.d@gmail.com}
\author{P.\ Mandal$^{2}$}
\author{A.\ Sundaresan$^{2}$}
\author{N.\ S.\ Vidhyadhiraja$^{1}$}\email{raja@jncasr.ac.in}
\affiliation{$^{1}$Theoretical Sciences Unit, Jawaharlal Nehru Centre For
Advanced Scientific Research, Jakkur, Bangalore 560064, India.}
\affiliation{$^{2}$Chemistry and Physics of Materials Unit, Jawaharlal Nehru Centre For Advanced Scientific Research, Jakkur, Bangalore 560064, India.}
\begin{abstract}
We present combined experimental and theoretical studies on the magnetic properties of a solid solution between yttrium orthoferrite and yttrium orthochromite systems, YFe$_{1-x}$Cr$_x$O$_3$ (0 $\leq$ x $\leq$ 1) where Fe$^{3+}$ and Cr$^{3+}$ ions are distributed randomly at the same crystallographic site (4b). We found that all the compositions exhibit weak ferromagnetism below the N\'eel temperature that decreases non-linearly with increasing $x$, while certain intermediate compositions ($x = 0.4,0.5$) show a compensation point and magnetization reversal. This unusual behavior is explained based on a simple model comprising the isotropic superexchange and the antisymmetric Dzyaloshinskii-Moriya interactions. This model explains the magnetization behavior in the entire range of doping and temperature including the magnetization reversal which results from an interplay of various DM interactions such as, Fe-O-Fe, Cr-O-Cr and Fe-O-Cr.
\end{abstract}

\maketitle

\section{Introduction}
Rare-earth orthoferrites  and orthochromites with the general formula RMO$_3$, where R = Rare earth or Yttrium and M = Fe and Cr, have perovskite structure with orthorhombic distortion (space group: Pnma) and an antiferromagnetic ground state.  Below the N\'eel temperature $T_N$, all these compounds exhibit a weak ferromagnetic behavior, arising from a slight canting of the antiferromagnetic backbone, similar to that observed in compounds such as $\alpha$-Fe$_2$O$_3$ and few transition metal carbonates. Such weak ferromagnetism (WFM) could result from either an antisymmetric superexchange, also known as Dzyaloshinskii-Moriya (DM) interaction or single-ion magnetic anisotropy or both\cite{PhysRev.120.91,PhysRev.125.1843}. In  orthoferrites and orthochromites, although both of these mechanisms operate,  it has been argued that the antisymmetric exchange interaction is the dominant cause of the observed WFM\cite{PhysRev.125.1843}. Neutron diffraction studies have shown that the magnetic structure is G-type with the magnetic easy axis lying along the $z$-direction\cite{Koehler1957100,Judin1966554}. In these weakly ferromagnetic orthoferrites, the spontaneous moment orients along the $y$-direction whereas the Dzyaloshinskii vector ${\mathbf D}$ points along the $x$-direction\cite{PhysRev.125.1843,1.1735136}. When R is a magnetic ion, most of these compounds undergo a spin re-orientation transition below which the direction of easy axis is changed to $y$. At low temperatures, some compounds exhibit magnetization reversal (MR) due to antiferromagnetic coupling of R-moments with the canted Cr-moments\cite{Yoshii2001204,Yoshii200184,Khomc}.

In a similar orthorhombic compound with nonmagnetic R-ion, namely YVO$_3$, weak ferromagnetism and MR have been reported\cite{YVO3}. The origin of MR has been explained based on a competition between DM interaction and single-ion magnetic anisotropy\cite{PhysRevB.62.6577}(SIMA). MR is also well known in several ferrimagnetic systems such as spinels\cite{PhysRev.90.487.2,PhysRevLett.4.119}, garnets\cite{1.1723094} and  Prussian blue analogs\cite{PhysRevB.56.11642,PhysRevLett.82.1285}. In these materials, MR has been explained by different temperature dependence of sublattice magnetization arising from different crystallographic sites, as predicted by N\'eel for spinel systems. In antiferromagnetic perovskite systems, like YVO$_3$, the magnetic ions occupy a single crystallographic site and therefore N\'eel's mechanism cannot explain the MR\cite{PhysRevB.82.100416,Mandal2013408}. In previous studies we demonstrated temperature induced MR in several canted antiferromagnetic oxides with nonmagnetic R-ion and mixed transition metal ions such as La$_{{1-x}/2}$Bi$_{x/2}$(Fe$_{0.5}$Cr$_{0.5}$)O$_3$, BiFe$_{0.5}$Mn$_{0.5}$O$_3$ and YFe$_{1-x}$Mn$_x$O$_3$ (0.1$\leq$x$\leq$0.45)\cite{PhysRevB.82.100416,Mandal2013408,0953-8984-21-48-486002}. In these systems, magnetic ions (Fe, Cr and Mn) in trivalent state are disordered at the B-site of the perovskite.

Here, we report a systematic and combined, experimental and theoretical study of a solid solution of two weakly ferromagnetic materials namely YFeO$_3$ and YCrO$_3$, that have
$T_N\simeq 640 K$ and $140 K$ respectively. We find, predictably, that at low and high values of $x$ in YFe$_{1-x}$Cr$_x$O$_3$, the compounds show WFM behaviour. However,
for intermediate compositions $x=0.4$ and $0.5$, a surprising temperature-dependent MR is observed. The possibility of finding MR in this system was indeed conjectured more than three decades
ago\cite{Kadomtseva}, but was not demonstrated until recently~\cite{1.3590714} (for $x=0.5$). The previously mentioned mechanisms for MR do not explain our findings. Thus, based on the experimental results and previous theoretical insights, we propose a simple mechanism based on an interplay of competing DM interactions that is able to explain the data quantitatively.

Polycrystalline samples of YFe$_{1-x}$Cr$_x$O$_3$ (0$\leq$x$\leq$1) have been synthesized by solid state reaction route by mixing stoichiometric amount of Y$_2$O$_3$ (preheated at 1223 K), Fe$_2$O$_3$ and Cr$_2$O$_3$ and sintered at 1743 K for 24 hours with several intermittent grindings. Rietveld refinement was carried out on the room temperature x-ray powder diffraction (XRPD) data collected with Bruker D8-Advance diffractometer. Magnetic measurements were carried out with a vibrating sample magnetometer in a Physical Property Measurement System (PPMS), Quantum Design, USA.

\begin{figure}[h!]
\centerline{\includegraphics[angle=0,width=1.0\columnwidth]{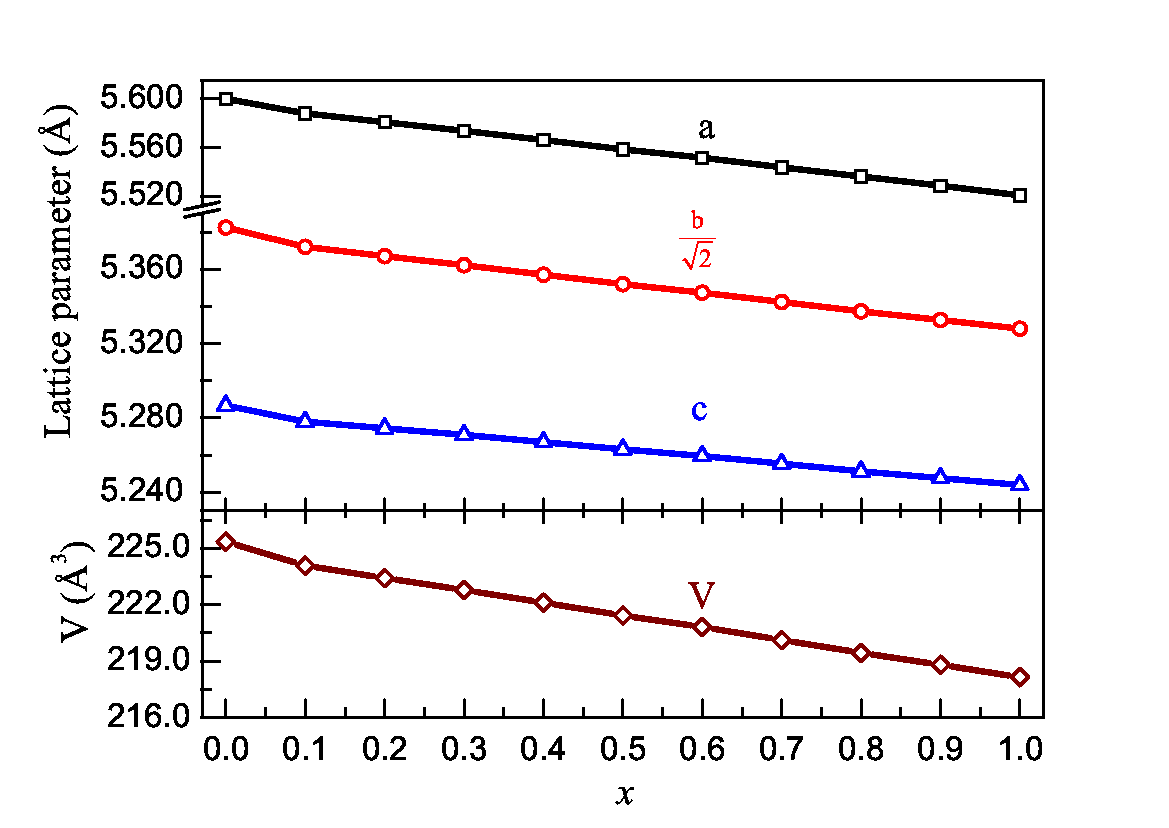}}
\caption{(color online) Variation of lattice parameters and volume as a function of
$x$ in YFe$_{1-x}$Cr$_x$O$_3$.}
\label{fig:fig1}
\end{figure}
A complete solid solution exists in YFe$_{1-x}$Cr$_x$O$_3$ as the two end members YFeO$_3$ and YCrO$_3$ have
 the orthorhombic structure (Pnma). In this structure, the Fe and Cr ions are randomly distributed at the 4b site. Unit cell parameters $a, b, c$ and cell volume $V$, as extracted from Rietveld refinement, are shown in figure~\ref{fig:fig1}. All these parameters decrease with $x$ and follow Vegard's law as expected from the difference in ionic radii between Fe$^{3+}$ and Cr$^{3+}$ (0.645 and 0.615 {\AA}), respectively. Field-cooled magnetization measurements of YFe$_{1-x}$Cr$_x$O$_3$ at an applied magnetic field of 100 Oe for various $x$ in the temperature interval 10 - 650 K were carried out. Our magnetization measurements reveal that the N\'eel temperature varies non-linearly with $x$. Further, all samples exhibit weak ferromagnetism below T$_N$ and the samples with $x$ = 0.4, and 0.5 exhibit magnetization reversal. These results are explained based on the model described below.

Neutron scattering results show that the Y(Fe,Cr)O$_3$ system is a G-type N\'eel antiferromagnet for all doping. As the Fe and Cr ions are disordered at the B-site, the Hamiltonian must have antiferromagnetic superexchange interaction term, of the form $J_{ij} {\hat{\textbf{S}_i}}\cdot {\hat{\textbf{S}_j}}$ with $J_{ij}>0$ for three possible pairs\cite{1.3590714}, namely Fe-O-Fe, Cr-O-Cr and Fe-O-Cr. The Fe$^{3+}$ ions have a spin $S=5/2$ while the Cr$^{3+}$ have $S=3/2$. In a solid solution, the two ions would occupy sites randomly with probability $P_{Fe}=(1-x)$ and $P_{Cr}=x$. The superexchange term will only be able to explain the antiferromagnetic order, while the explanation of weak ferromagnetism requires invoking other mechanisms such as Dzyaloshinskii-Moriya interactions (DM) or the single-ion magnetic anisotropy (SIMA). In the parent compounds, YFeO$_3$ and YCrO$_3$, the WFM has been understood as arising purely from DM interactions of the Fe-Fe and Cr-Cr pairs respectively. For compositions not equal to 0 or 1, we continue to keep only DM interactions, neglecting SIMA completely. Furthermore, for $x\neq 0,1$, we must consider $D_{FeCr}$ along with the usual $D_{FeFe}$ and $D_{CrCr}$ interactions. In previous work on this system, the authors have suggested~\cite{1.3590714}the choice of DM interactions to have the form $\vec{D}_{FeFe}\;||\;\vec{D}_{CrCr}\;||\; -\vec{D}_{FeCr}$. We arrive at the same conclusion by eliminating other possibilities, and indeed find excellent explication of the experimental results with this approach (described below).

\section{Model and Results}

The Heisenberg hamiltonian\cite{PhysRev.115.2} without the DM or SIMA can be written in a general form as
\begin{align}
{\cal H}_s=\sum_{ij\alpha\beta}J^{ij}_{\alpha\beta}{\hat{{\textbf{S}}}}_{i\alpha}
\cdot\hat{{\textbf{S}}}_{j\beta}\,.
\end{align}
Here $i,j$ denote lattice site indices and $\alpha,\beta$ indicate the type of magnetic ion, which in our case could be either Fe or Cr.
Employing the molecular field approximation (MFA) for the antiferromagnetic case and ignoring the spin-flip terms, the above Hamiltonian reduces to
\begin{equation}
{\cal H}_{MFA}=-\sum_{\alpha=Fe,Cr} (H^{eff}_{B\alpha}\sum_{i\in A}
\hat{S}_{i\alpha}^z+ H^{eff}_{A\alpha}\sum_{i\in B}
 \hat{S}_{i\alpha}^z)
\end{equation}
with $z$ being the number of nearest neighbours and $H^{eff}_{B\alpha}$ is the mean field due to the B-sublattice atoms on the A-sublattice which is explicitly given by
\begin{equation}
 H^{eff}_{B\alpha}=2zP_{Fe} J_{\alpha Fe} \langle\hat{S}_{Fe}^z
\rangle_B+ 2z P_{Cr}J_{\alpha Cr} \langle\hat{S}_{Cr}^z\rangle_B\,.
\end{equation}
Correspondingly, $H^{eff}_{A\alpha}$ is the mean field due to the A-sublattice atoms on the $\alpha$-atoms in the B-sublattice.
Note that the mean fields are different for the Fe and the Cr atoms and depend on doping levels as well. Using the above MFA Hamiltonian, the partition function may be obtained in a straightforward way by tracing over the $\hat{S}^z_{Fe}$ and $\hat{S}^z_{Cr}$ eigenvalues, which yields the self-consistent equations that describe the temperature dependence of the Fe and Cr spins as
$\langle \hat{S}^z_{\alpha}\rangle = S_{\alpha} B_J(X_{\alpha})\;\;\;\alpha=Fe, Cr\,, $
where the sublattice index has been suppressed (for clarity) and
$B_J(x)$ is the Brillouin function; The $X_\alpha$'s are given by
\begin{equation*}
X_{Fe} = \frac{2zS_{Fe}}{kT} \left[J_{Fe,Fe} P_{Fe}^2 \langle
\hat{S}^z_{Fe}\rangle+J_{Fe,Cr}\,P_{Fe}P_{Cr} \langle \hat{S}^z_{Cr}\rangle \right]
\end{equation*}
and
\begin{equation*}
X_{Cr} =\frac{2zS_{Cr}}{kT} \left[zJ_{Cr,Fe}\, P_{Cr}P_{Fe} \langle
\hat{S}^z_{Fe}\rangle +J_{Cr,Cr}\,P_{Cr}^2 \langle \hat{S}^z_{Cr}\rangle\right]\,.
\end{equation*}
Solving the above coupled nonlinear equations, we can obtain the A-sublattice magnetization as
$M_A(T)=\frac{n}{2}g\mu_B(P_{Fe}\langle \hat{S}^z_{Fe}\rangle_A + P_{Cr}\langle \hat{S}^z_{Cr}\rangle_A) \,.$
For a perfect antiferromagnet considered until now, the total magnetization ($M_{tot}=M_A+M_B$) is naturally zero. To find the N\'eel temperature of the doped system, we can follow the usual procedure of linearizing the above equations in the limit $T\rightarrow T_N^-$, where we expect $\langle S^z_{\alpha}\rangle \rightarrow 0$. It is easy to see that the requirement of getting non-zero values of $\langle S^z_{\alpha}\rangle$ yields a $4\times 4$ determinant which when further simplified yields the equation
$1-2LM-2KMLN-N^2-K^2+L^2M^2+K^2N^2=0\,$
where
\begin{align}
K&= \frac{2zP_{Fe}^2 S_{Fe}(S_{Fe}+1)}{3KT}J_{FeFe}\,,\hspace{0.3cm} L=\frac{P_{Cr}}{P_{Fe}}\frac{J_{FeCr}}{J_{FeFe}}\,K
\nonumber \\
N &= \frac{2zP_{Cr}^2 S_{Cr}(S_{Cr}+1)}{3KT}J_{CrCr}\,,\hspace{0.3cm} M=\frac{P_{Fe}}{P_{Cr}}\frac{J_{FeCr}}{J_{CrCr}}\,N\,. \nonumber
\end{align}

\begin{figure}[h!]
\centerline{\includegraphics[angle=0,width=1.0\columnwidth]{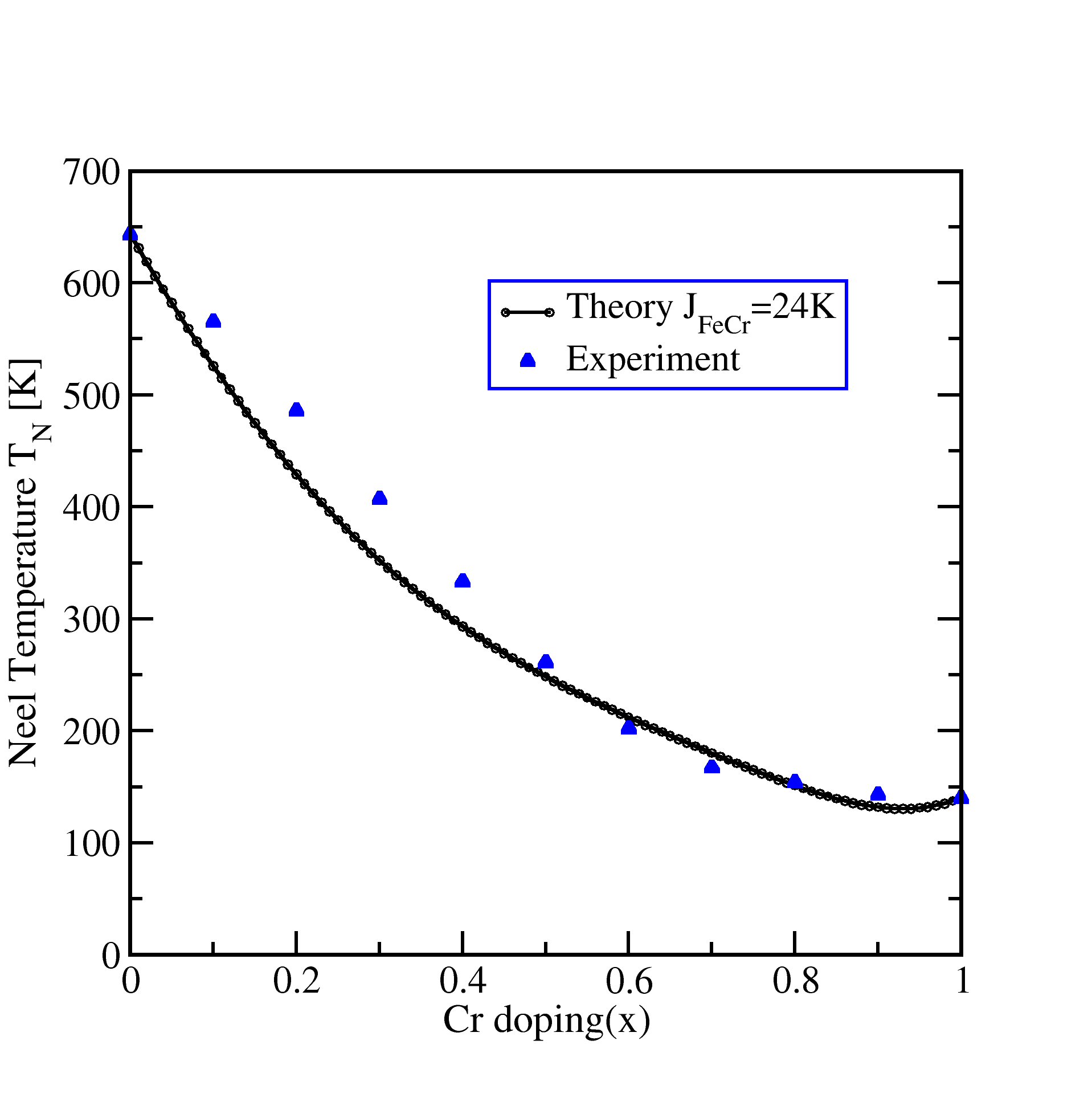}}
\caption{(color online) Variation of N\'eel temperature (experiment:triangles and theory:circles)
with increasing Cr content. The theoretical curve has been computed with
$J_{FeCr}$ = 24 K.}
\label{fig:TN}
\end{figure}

For $J/kT \ll 1$, we retain terms of ${\cal{O}}((J/kT)^2)$ and neglect the higher order terms, thus getting
\begin{equation}
T_N(x)  = \frac{2z}{3k}
\left(\sum_{\alpha\beta}S_{\alpha}(S_\alpha+1)S_\beta(S_\beta+1)
P_\alpha^2P_\beta^2 J_{\alpha\beta}^2 \right)^\frac{1}{2}
\end{equation}
where $\alpha,\beta=Fe,Cr$, $P_\alpha=(1-x)\delta_{Fe,\alpha} + x\delta_{Cr,\alpha}$
is the probability of site occupancy, and the spins are given by $S_{Fe}=5/2$ and $S_{Cr}=3/2$. The nearest neighbour coordination number $z$ is 6 for the present case. In a previous molecular field theoretical study of the doped system, the $T_N$ {\it vs} $x$ expression was obtained\cite{JPSJ.18.1140}, which was different than the one obtained above. However, their result seems inconsistent with undoped system N\'eel temperature expression, i.e., if $x=0$ or $x=1$, we should recover the pure compound N\'eel temperature expressions, which their expression does not while the above equation does. This inconsistency could be because they neglected to consider the probabilistic aspect of the occupancy of the site on which the mean-field is acting. Using the above equation for $x=0$ and $x=1$ with $T_N$ from experimental measurement~\cite{Judin1966554,PhysRevB.72.220101} being 640 K and 140 K respectively, we can extract the values of $J_{FeFe}$ and $J_{CrCr}$ as 18.4 K and 9.3 K respectively. These small $J/kT (\sim 0.03-0.07)$ values self-consistently justify the neglect of cubic and higher order terms in $(J/kT)$. To find $J_{FeCr}$, we carry out a best fit of the above equation to the experimentally measured $T_N(x)$, as shown in figure~\ref{fig:TN} (circles: theory and experiment: diamonds). This yields a $J_{FeCr}$ = 24.0 K, which is surprisingly higher than the
superexchange in the parent compounds. The agreement of experimental data with the theoretical expression given above is remarkable.
The dominant $x$ dependence near low $Cr$ concentration is $T_N(x\rightarrow 0) = T_{N,Fe} (1-2x)$ and at high concentration, close to YCrO$_3$ is $T_N(x\rightarrow 1) = T_{N,Cr} (1-2(1-x))$. The N\'eel temperatures at intermediate concentrations are, as usual, somewhere between those of the two parent compounds, but at the two ends, as is seen in experiment as well, the doped compound has a lower $T_N$ than the parent compound.

Now we build upon the underlying antiferromagnetism and outline our approach for understanding the weak ferromagnetism and magnetization reversal in the YFe$_{1-x}$Cr$_x$O$_3$ system. To begin with, consider the parent compounds, YFeO$_3$ and YCrO$_3$. As the experimental results(see later) show, the parent compounds are weak ferromagnets~\cite{Judin1966554,PhysRevB.72.220101}. Such weak ferromagnetism (WFM) is caused by a small canting of the underlying antiferromagnetic lattice. In general, the canting could be strongly temperature dependent and could arise due to a variety of reasons most important of which are the DM and the SIMA interactions. For YVO$_3$, it was argued~\cite{PhysRevB.62.6577} that a strongly temperature dependent DM interaction energy along with a staggered easy axis leads to a magnetization sign reversal with decreasing temperature. The authors did include SIMA in their semi-classical approach, albeit without temperature dependence. Although a good agreement with experimental data was achieved, the anisotropy term was found (by fitting to experiments) to be about 1.7 times the Heisenberg exchange. Such a result violates the initial assumption that the DM/SIMA interactions are much weaker compared to the Heisenberg exchange and may thus be treated perturbatively. Furthermore, such a large anisotropy is physically unjustified.

   We present a alternative approach to the present problem that is minimalist but physically and internally consistent. It has been argued in the literature (see for instance ~\cite{PhysRev.120.91}) that for relatively high Neel temperatures ($T_N\gtrsim$ 100 K), the canting is caused primarily by DM interactions, while for lower $T_N$ materials, the single-ion anisotropy dominates. Since the parent compounds have high $T_N$, our minimalist approach neglects the effects of SIMA completely and aims to understand all of the magnetization behaviour in the doped system purely through DM interactions.

The Hamiltonian including only the superexchange and the DM interactions in the absence of an external field is given by
\begin{align}
{\cal H_S} = \sum_{\langle ij\rangle} ( J_{ij} \hat{\textbf{S}}_i\cdot \hat{\textbf{S}}_j -
 \vec{D}_{ij}\cdot (\hat{\textbf{S}}_i\times \hat{\textbf{S}}_j))\,.
\end{align}
The classic DM interaction has been derived by Dzyaloshinskii and Moriya~\cite{PhysRev.120.91} for the non-centrosymmetric anion mediated antisymmetric exchange interaction between two {\em same} magnetic ions. Yamaguchi extended
this result to different kinds of magnetic ions,~\cite{Yamaguchi1974479} namely R$^{3+}$ and Fe$^{3+}$ in RFeO$_3$. Extending this idea to the doped system here, we consider DM interactions between neighbouring $Fe$ and $Cr$ ions.
\begin{figure}[h!]
\centerline{\includegraphics[angle=0,width=0.7\columnwidth]{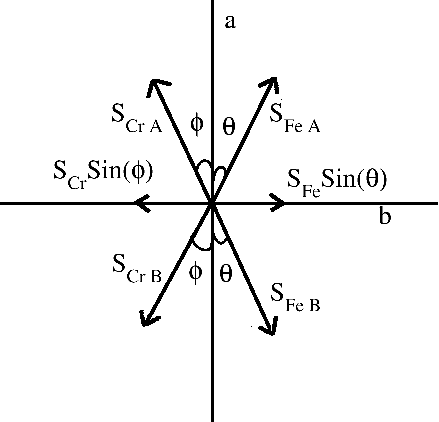}}
\caption{Schematic spin structure for the YFe$_{1-x}$Cr$_x$O$_3$ system.}
\label{fig:model_FeCr}
\end{figure}
We employ the molecular field approximation again, and with the model shown in figure~\ref{fig:model_FeCr},
the average energy reduces to a function of the canting angles $\theta$ and $\phi$.
To find the canting angles, we minimize the energy above with respect to $\theta$ and $\phi$. In the small angle limit,
we end up with two linear equations,
which are in terms of $J_{FeFe}, J_{CrCr}, J_{FeCr}, D_{FeFe}, D_{CrCr}$ and $D_{FeCr}$.

The superexchange parameters $J_{\alpha\beta}$ are obtained from the experimental N\'eel temperatures and the $D_{\alpha\beta}$ are obtained by comparing theory to the experimentally measured weak ferromagnetic magnetization in the parent and the doped compounds. For the parent compounds ($x=0$ and $x=1$), we find temperature independent canting angles (as in ~\cite{PhysRevLett.5.13})
$\theta=D_{FeFe}/2J_{FeFe}\;\;{\rm and}\;\;\phi=D_{CrCr}/2J_{CrCr}$.
The net magnetization is given in the general case (for a powder sample\cite{LimaJr2008622}) as
\begin{equation}
M_{net}= \frac{ng\mu_B}{2}\left(P_{Fe}\langle \hat{S}^z_{Fe}\rangle \theta + P_{Cr}
\langle \hat{S}^z_{Cr}\rangle\phi\right)
\end{equation}
where the average $z$-component of each of the spins is given by the earlier found mean field expressions. By comparing the temperature dependence  of the theoretically obtained magnetization to the experimentally measured one for the parent compounds, we can extract the magnitudes of the Dzyaloshinskii vectors, $D_{FeFe}$ and $D_{CrCr}$.

\begin{figure}[h!]
\centerline{\includegraphics[angle=0,width=1.0\columnwidth]{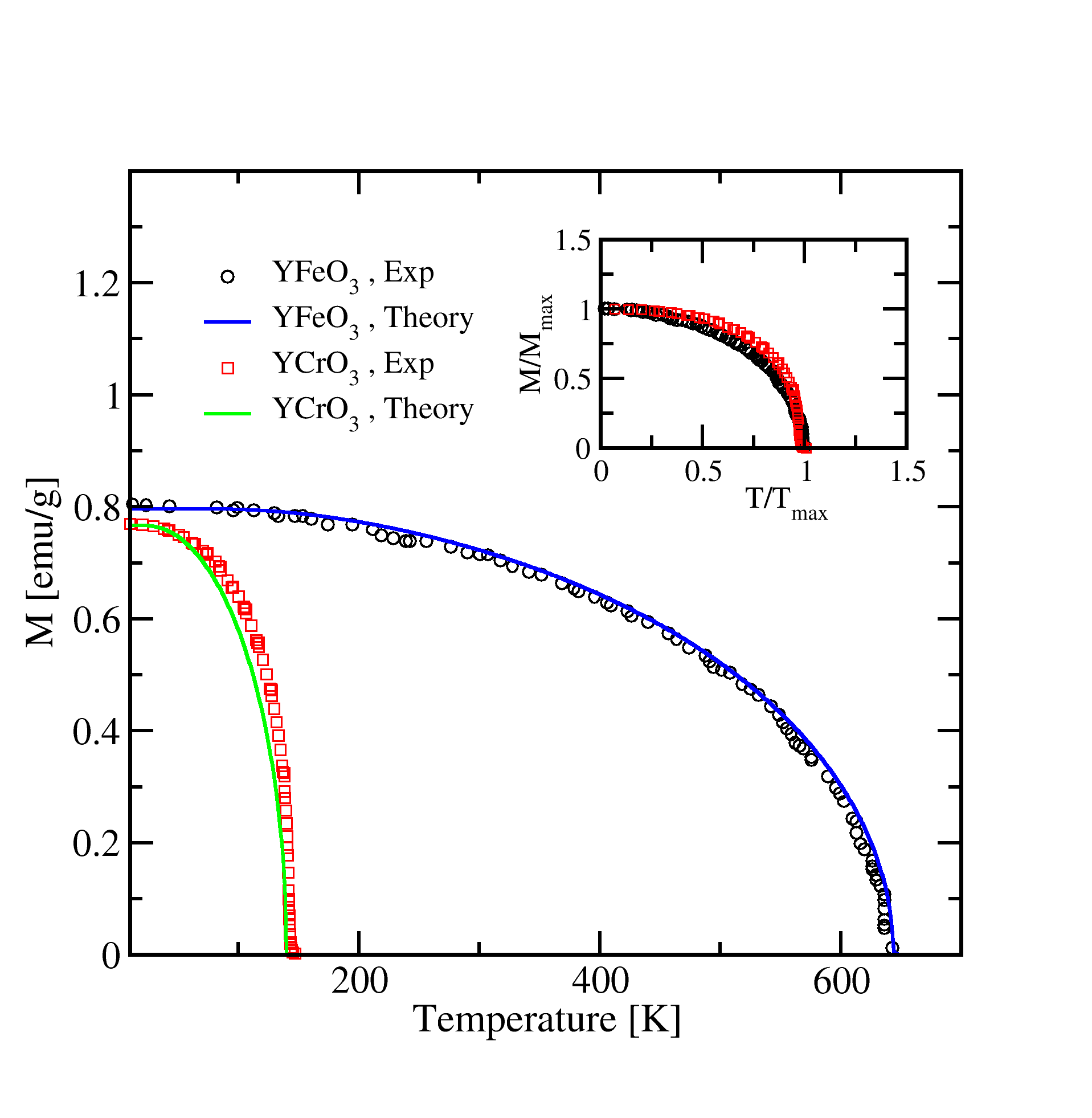}}
\caption{(color online) Magnetization (experiment and theory, see legends)  as a function of temperature for the parent compounds YFeO$_3$ and YCrO$_3$.  The inset shows that a common mechanism underlies the WFM of both the parent compounds, since the $M/M_{\rm max}$ {\it vs.} $T/T_N$ is almost identical for both.
}
\label{fig:parent}
\end{figure}

Taking the physically reasonable~\cite{Judin1966554} DM values of $D_{FeFe}$ and $D_{CrCr}$ as 0.4K and 0.32K respectively, we compute the magnetization {\it vs.} temperature.
As shown in figure~\ref{fig:parent}, the description of weak ferromagnetism purely using DM interactions agrees remarkably with experiment. The inset shows that a common mechanism underlies the WFM of both the parent compounds, since the $M/M_{\rm max}$ {\it vs.} $T/T_N$ is almost identical for both. The slight deviation of theory from experiment for YCrO$_3$ suggests that single-ion magnetic anisotropy might need to be invoked to get a better fit.
The canting angles $\theta$ and $\phi$ do not depend on temperature in the parent compounds. We will see below that this will not be the case for the doped case.

\begin{figure}[h!]
\centerline{\includegraphics[angle=0,width=1.0\columnwidth]{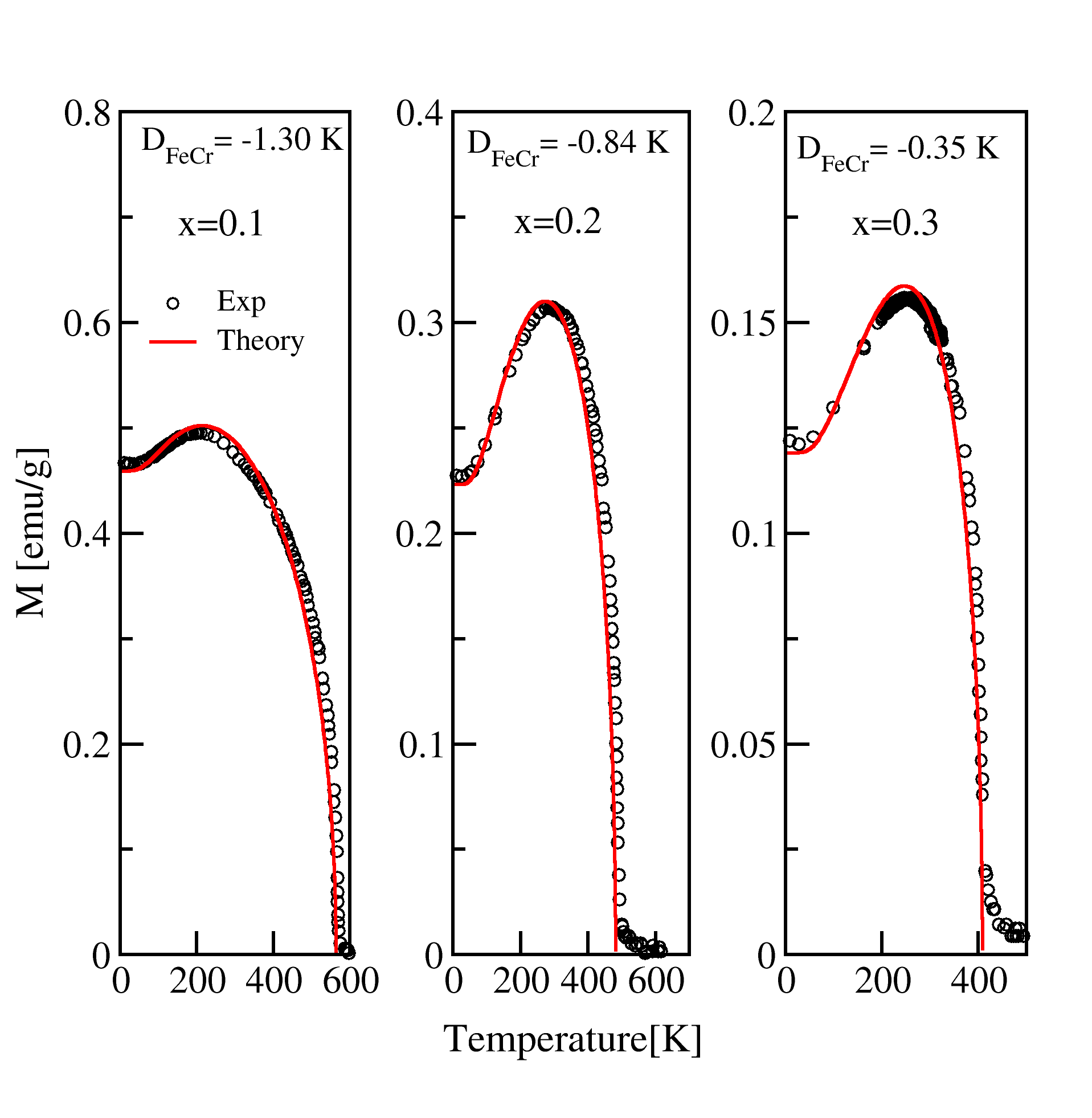}}
\caption{(color online) Magnetization (experiment:black and theory:red) as a function of temperature for three compositions, $x=0.1$ (left), $0.2$(middle) and $0.3$ (right).}
\label{fig:123}
\end{figure}

The DM interaction between the Fe and Cr atoms must be expected to depend on the concentration $x$. So, to obtain the values of $D_{FeCr}$ as a function of $x$, we follow the
same route as for the parent compounds.  The parameter $D_{FeCr}$ is obtained for each concentration $x$ using a best fit to the experimental data. The g factor has been varied slightly for obtaining a quantitative fit, which amounts to a simple multiplicative scaling of the $y$-axis. We first consider the doping range $x = 0.1, 0.2$ and $0.3$. The experimental data is shown in figure~\ref{fig:123} (black circles). The N\'eel temperature decreases with increasing $x$, and the overall magnetization value also comes down.  A broad maximum appears and this is a characteristic signature of spin reorientation. The limiting zero temperature ($T\rightarrow 0$)  magnetization is seen to decrease sharply. Thus it can be expected that at higher doping, a magnetization reversal will occur, and indeed this is seen as we show below. Before that, let us discuss the comparison to theory. In the top panels of figure~\ref{fig:123}, the theoretically computed magnetization (in red) with the same exchange couplings as before and best fit values of $D_{FeCr}$=-1.3 K, -0.84 K and -0.35 K for $x$=0.1,0.2,0.3 respectively are superimposed on the experimental data. The agreement is seen to be excellent over the entire temperature range. The canting angles, as inferred from the above comparison (not shown) depend on temperature and in fact increase monotonically in magnitude. Thus the different dependences of $\theta$ and $\phi$ on $T$ seems to be responsible for the continuous spin reorientation.

\begin{figure}[h!]
\centering{
\includegraphics[angle=0,width=\columnwidth]{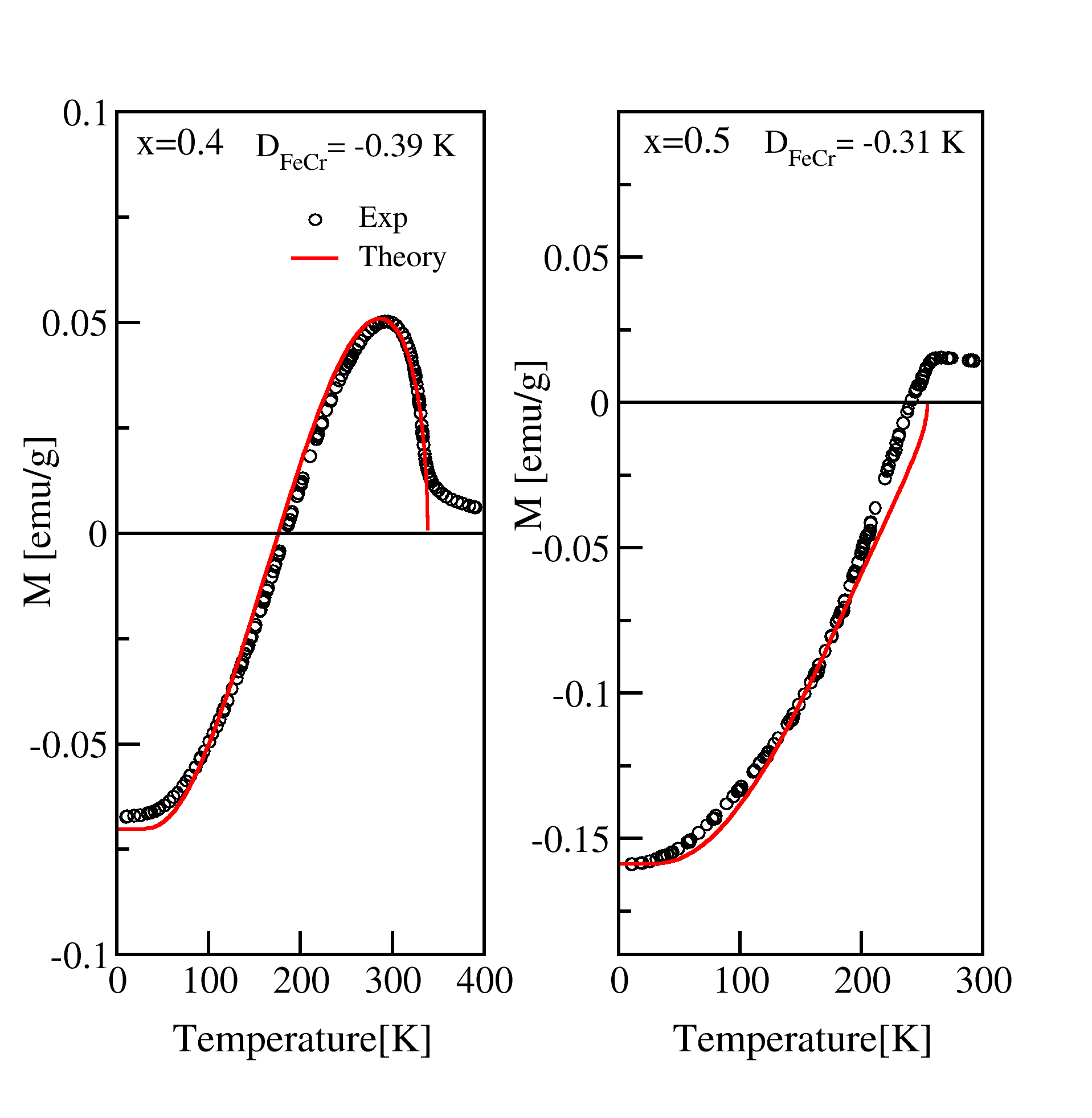}
}
\caption{(color online) Magnetization (experiment:black and theory:red) as a function of temperature for $x=0.4$ (left) and $0.5$ (right). Magnetization reversal is seen in this composition range. }
\label{fig:45}
\end{figure}

The experimental data for $x=0.4$ and $0.5$ is shown (in black circles) in figure~\ref{fig:45}. For $x=0.4$, a smooth magnetization reversal with a compensation point at $T_{\rm comp}\sim $ 170 K is seen. In fact, for temperatures below the compensation point, the magnetic behaviour must be described as weak diamagnetism, since these are field cooled experiments, albeit with a small applied field (100 Oe). For $x=0.5$, the onset of antiferromagnetism also signals WFM, but with a slight decrease in temperature, magnetization reversal occurs. The theory curves are again superimposed (in red) onto the experiment, with $D_{FeCr}$=-0.39 K and -0.31 K and again the whole temperature dependence is captured faithfully. Thus, in this approach, the magnetization reversal may be argued to arise from the competition between the magnetization of the Fe-Fe and Cr-Cr pairs {\it vs.} that of the Fe-Cr pairs, induced by the interatom DM interaction. In other words, if $D_{FeCr}$ were zero, then the magnetization of Fe atoms and the Cr atoms would just add up, and there would be no temperature dependent reversal or even spin reorientation. But in the presence of $D_{FeCr}$, which is antiparallel to $D_{FeFe}$ and $D_{CrCr}$, the Fe-Cr nearest neighbour pairs would exhibit WFM in a direction opposite to the Fe-Fe and Cr-Cr neighbour pairs, and thus at values of composition where heterogenous pairs are expected to be large in number as compared to homogenous pairs, one can expect a reversal of magnetization with decreasing temperature. The composition at which the reversal should occur should be determined by the relative magnitudes of the isotropic exchange strength. This is larger for Fe-Fe ($\sim $ 18 K) than for Cr-Cr ($\sim$ 9K), and hence the Cr atoms, which would normally order antiferromagnetically at much lower temperatures, begin ordering much above the N\'eel temperature of the parent compound YCrO$_3$, because of the $J_{FeCr}$ exchange. Thus the reversal must happen closer to YFeO$_3$ and indeed it is seen at $x$=0.4.

\begin{figure}[h!]
\centering{
\includegraphics[angle=0,width=1.0\columnwidth]{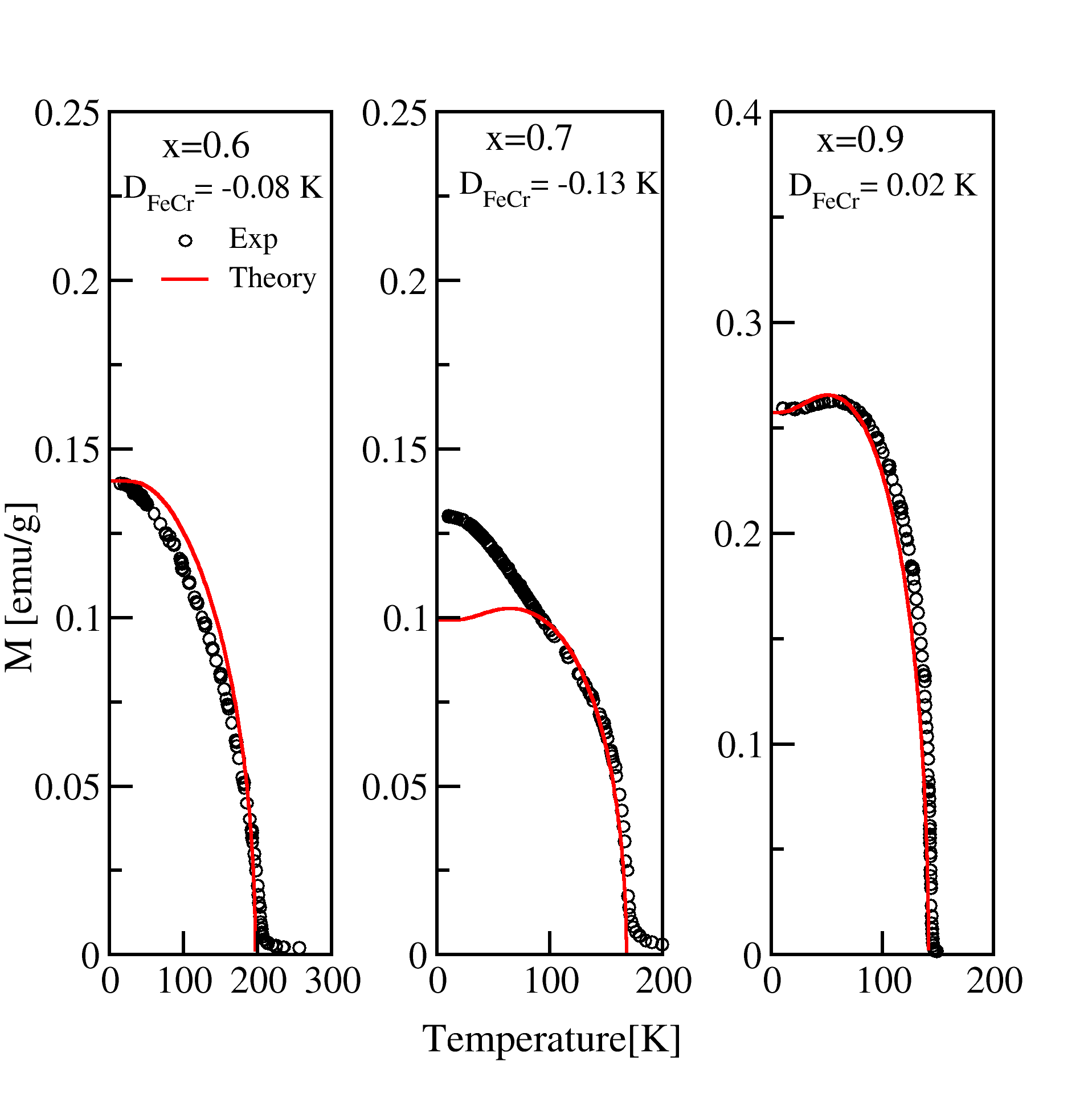}
}
\caption{(color online) Temperature dependent magnetization
(experiment: black and theory:red)
for $x=0.6$ (left), $0.7$ (middle) and $0.9$(right).}
\label{fig:679}
\end{figure}

The compositions $x=0.6, 0.7$ and $0.9$, which are closer to the parent $Cr$ compound YCrO$_3$ are discussed in figure~\ref{fig:679}.  It is seen that WFM is recovered for these compositions, since the number of homogenous pairs (Cr-Cr) would be expected to be larger than the heterogenous pairs. The theory agrees reasonably well with experiment. For $x=0.7$, the middle panel of figure~\ref{fig:679} shows that the agreement between theory and experiment is excellent for temperatures $\gtrsim$ 100 K, while at low temperatures, the theory predicts lower magnetization than what is observed in the experiment.
We conjecture that at higher concentrations of $Cr$, the theory might need to be modified and other interactions like the single-ion-anisotropy that have been neglected in the present approach might have to be included to get better agreement. In fact, for $x=0.8$, the experiments (not shown here) show two magnetization reversals, but the absolute value of magnetization is very small and almost comparable to the field induced magnetization values. Such a double reversal simply cannot be captured by the present theoretical approach.

\section{Conclusions}

\begin{figure}[h!]
\centering{
\includegraphics[angle=0,width=1.0\columnwidth]{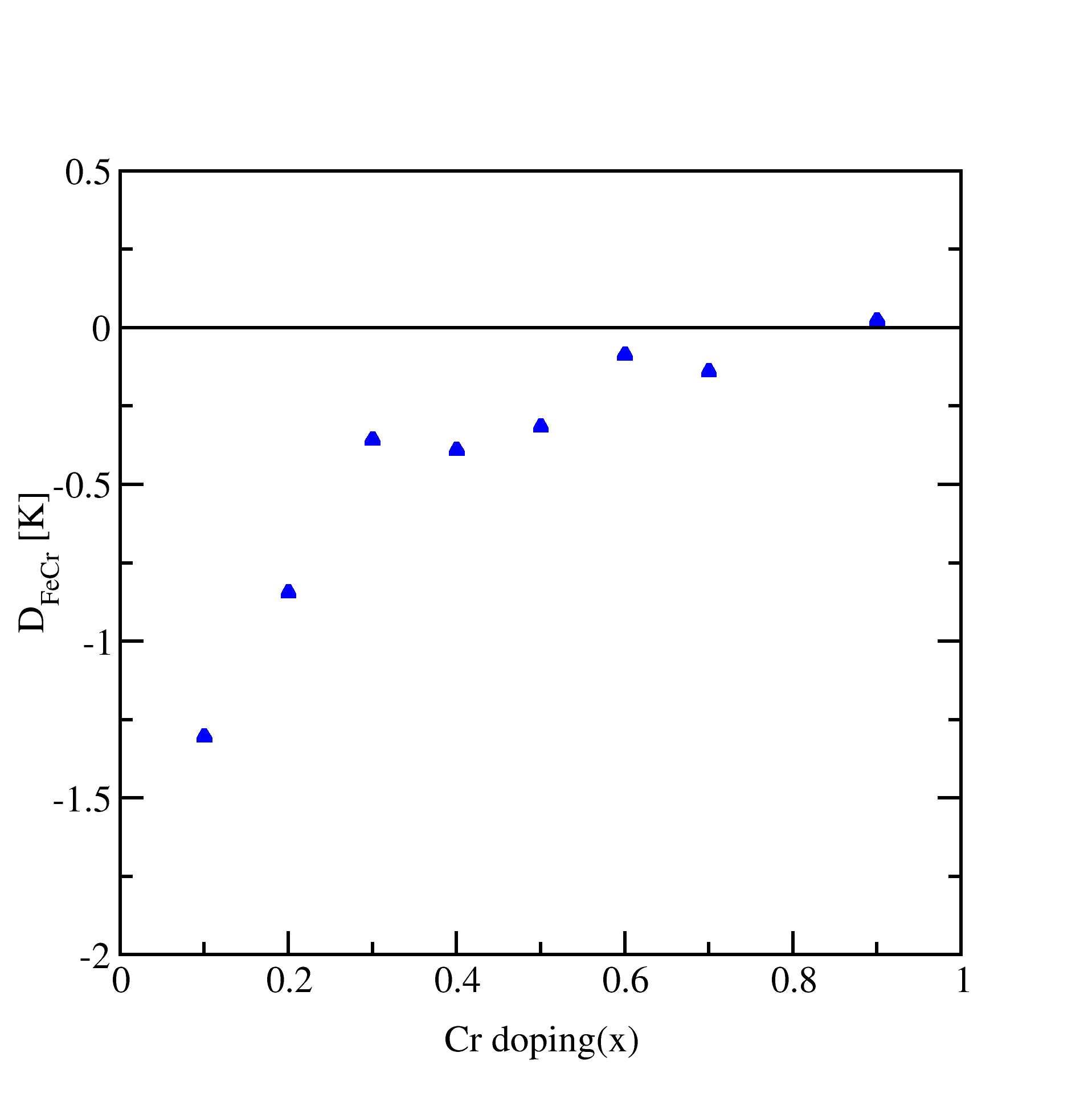}
}
\caption{(color online) The DM interaction $D_{FeCr}$ between the Fe and Cr atoms, as obtained from the comparison of theory and experiment is shown as a function of composition $x$. It is intriguing to note that it is maximum in magnitude close to YFeO$_3$ and decreases almost monotonically with increasing $x$.}
\label{fig:dmx}
\end{figure}

In conclusion, we have investigated the magnetization behaviour as a function of temperature and doping for the solid solution YFe$_{1-x}$Cr$_x$O$_3$. An interplay of different DM interactions leads to interesting spin-reorientation and magnetization reversal.
It is interesting to note that even though the parent compounds are weak ferromagnets with monotonic temperature dependence of magnetization $M(T)$, the mixed compounds display magnetization reversal and a non-monotonic $M(T)$. In our approach, this behaviour finds a natural explanation in terms of the doping dependence of $D_{FeCr}$ (shown in figure~\ref{fig:dmx}) and the negative sign of the parameter, which suggests that the DM vector ${\mathbf{D}}_{FeCr}$ is opposite in direction to the ${\mathbf{D}}_{FeFe}$ and ${\mathbf{D}}_{CrCr}$ vectors in the parent compounds.
This opposite direction introduces a competition between the canting driven by the DM interactions of the heterogenous pairs (Fe-Cr) versus the homogenous pairs (Fe-Fe and Cr-Cr). Thus we are able to obtain quantitative agreement between theory and agreement for the whole range of doping and temperature with a very simple, consistent and transparent approach. This also suggests that doping could be used very effectively to tune the antisymmetric exchange parameter. Furthermore, we opine that the interplay of various interactions considered here
must exist in other similar B-site disordered perovskite materials.

The canting angles were found to depend appreciably on temperature and doping. The DM interaction too seems to be dependent strongly on the composition. These two together suggest that changes in spin structure induce changes in the lattice structure, which implies the existence of significant spin-phonon coupling.  Indeed, recent experiments~\cite{1.3590714} have indicated a multiferroic nature of the YFe$_{0.5}$Cr$_{0.5}$O$_3$ material.  The microscopic justification for the $x$ dependence of the parameters obtained here would require a detailed analysis of the structural changes in the orthorhombic lattice due to the size differences in the Fe/Cr ions. Additionally, we would also require to find out the changes in the spin interactions due to the lattice distortions. These investigations will be the subject of future studies.

\section{Acknowledgments}

We thank CSIR and DST(India) for research funding. We would like to thank Prof. S. Ramasesha for very fruitful discussions. One of us (A.S) thank Quantum Design, USA for providing high temperature magnetization data for few samples.


\bibliographystyle{apsrev4-1}
\bibliography{apssamp}
\end{document}